\documentclass{aa}  
\usepackage{graphicx}
\usepackage{txfonts}
\usepackage{longtable}

\newcommand{\Teff}{\hbox{$T\sb{\rm eff}$}}          
\newcommand{\logg}{\hbox{$\log g$}}
\newcommand{\Msun}{\hbox{M$\sb{\odot}$}}
\newcommand{\Halpha}{\hbox{H$\alpha$}}

\begin{document}

\title{DB white dwarfs in the Sloan Digital Sky Survey Data Release 10
and 12}

   \author{D. Koester \inst{1} \and S.O. Kepler\inst{2}}

     \institute{Institut f\"ur Theoretische
     Physik und Astrophysik, Universit\"at Kiel, 24098 Kiel,
     Germany\\ 
     \email{koester@astrophysik.uni-kiel.de} 
     \and Instituto de Fisica, Universidade Federal do Rio Grande do Sul, 
     91501-900 Porto-Alegre, RS, Brazil
     }

   \date{Sep 21, 2015}

\abstract{} {White dwarfs with helium-dominated atmospheres (spectral types
  DO, DB) comprise approximately 20\% of all white dwarfs. There are
    fewer studies than of their hydrogen-rich counterparts (DA) and thus
  several questions remain open. Among these are the total masses and the
  origin of the hydrogen traces observed in a large number and the nature of
  the deficit of DBs in the range from 30\,000 - 45\,000\,K. We use the
  largest-ever sample (by a factor of 10) provided by the Sloan Digital Sky
  Survey (SDSS) to study these questions.}  {The photometric and spectroscopic
  data of 1107 helium-rich objects from the SDSS are analyzed using
  theoretical model atmospheres. Along with the effective temperature and
  surface gravity, we also determine hydrogen and calcium abundances or upper
  limits for all objects. The atmosphere models are extended with envelope
  calculations to determine the extent of the helium convection zones and thus
  the total amount of hydrogen and calcium present.}  {When accounting for
  problems in determining surface gravities at low \Teff, we find an average
  mass for helium-dominated white dwarfs of $0.606\pm0.004$\,\Msun, which 
    is very similar to the latest determinations for DAs. There are 32\% of
  the sample with detected hydrogen, but this increases to 75\% if only the
  objects with the highest signal-to-noise ratios are considered. In addition,
  10-12\% show traces of calcium, which must come from an external source. The
  interstellar medium (ISM) is ruled out by the fact that all polluted objects
  show a Ca/H ratio that is much larger than solar. We also present arguments
  that demonstrate that the hydrogen is very likely not accreted from the ISM
  but is the result of convective mixing of a residual thin hydrogen layer
  with the developing helium convection zone. It is very important to
  carefully consider the bias from observational selection effects when
  drawing these conclusions.}  {} 

   \maketitle

\section{Introduction}

Approximately 20\% of all white dwarfs have atmospheres dominated by
helium. Above \Teff $\approx$ 40\,000\,K, \ion{He}{ii} lines are strong and
the stars are classified as spectral type DO or DOA if, in addition, Balmer
lines of hydrogen are visible, or as DAO if these are dominant. Below this
temperature \ion{He}{i} lines are dominant and the spectral type is DB. If
traces of other elements are present, more letters are added to the type,
e.g., A for hydrogen, Z for metals \citep{Sion.Greenstein.ea83}. The existence
of almost pure hydrogen (DA) and almost pure helium atmospheres as a result of
gravitational settling is well understood, with the lightest element present
floating to the top of the outer layers \citep{Schatzman48}. Helium-rich white
dwarfs must have almost completely lost their outer hydrogen layer in a
previous evolutionary phase; the currently accepted scenario is the
born again or  late thermal pulse scenario \citep{Iben.Kaler.ea83}.

Questions remain as to the so-called DB gap and the origin of hydrogen traces
in a large number, perhaps the majority, of DB white dwarfs. The deficit of
DBs between 30\,000 and 45\,000\,K was first identified by
\cite{Liebert.Wesemael.ea86} and explained in terms of diffusion and
convective mixing by \cite{Fontaine.Wesemael87}.
\cite{Eisenstein.Liebert.ea06} and \cite{Kleinman.Kepler.ea13} found several
objects within the gap, but the number was still smaller than expected from
the simple cooling rates without changes of spectral types. In principle, the
observed traces of hydrogen could be explained by two competing theories:
accretion of interstellar matter (ISM) or convective mixing with a (thin)
hydrogen layer left over from the previous evolution (which, at the same time,
could also explain the DB gap).

Two recent studies have analyzed large samples of helium-rich white
dwarfs. \cite{Voss.Koester.ea07} used data for 71 objects from the Supernova
Ia Progenitor Survey \citep[SPY,][]{Napiwotzki.Christlieb.ea03}. The typical
resolution (depending on the seeing) was $\sim$0.4\,\AA, with the
signal-to-noise ratio (S/N) varying but $>15$. The study by
\citet[][henceforth BW11]{Bergeron.Wesemael.ea11} is the largest so far with
108 objects. The resolution from two different spectrographs was 3-6\,\AA,
with the S/N mostly above 50. In this work we use a sample of 1107 DB (and DBA
or DBZ) stars, thus increasing the sample more than tenfold the largest
previous ones. The resolution is $\sim$2.5\,\AA, the S/N ranges from 10 to 75
with an average of 20. The average quality of the spectra is inferior to the
BW11 sample in S/N and in resolution to the SPY sample. However, this is
compensated for by the huge size and homogeneity, which allows us to study
statistical properties of the helium-rich white dwarfs with unprecedented
quality.

\section{The sample}

Data Release 7 \citep[][DR7]{Kleinman.Kepler.ea13} of the Sloan Digital Sky
Survey (SDSS) contained spectra of 923 stars classified as DB (helium
atmospheres). DR10 \citep{Kepler.Pelisoli.ea15} added another 450 (including
subtypes DBA with hydrogen traces and DBZ with metals), and DR12 (Kepler et
al. 2015a) added 121 more, all of which are new
detections. The number of spectra is larger since several have two or three
spectra in the database. From this database we first selected all those with
S/N greater than 10.

After a first tentative fit with model spectra (see below for details) and a
visual inspection we eliminated all objects with peculiarities, such as a red
unresolved companion (DB+dM), presence of \ion{He}{ii} lines (DO), obvious
magnetic splitting of spectral lines, very marginal or invisible He lines
(DC), or spectra with strong artifacts. We did not eliminate DB stars with
traces of hydrogen (DBA) or calcium H+K lines (DBZ). This left us with a
sample of 1267 spectra of 1107 different objects, of which 13 had three and
136 had two spectra. In addition to the spectra, we used the SDSS photometry,
available for each object in the sample.

\begin{figure}
\centering 
\includegraphics[width=0.49\textwidth]{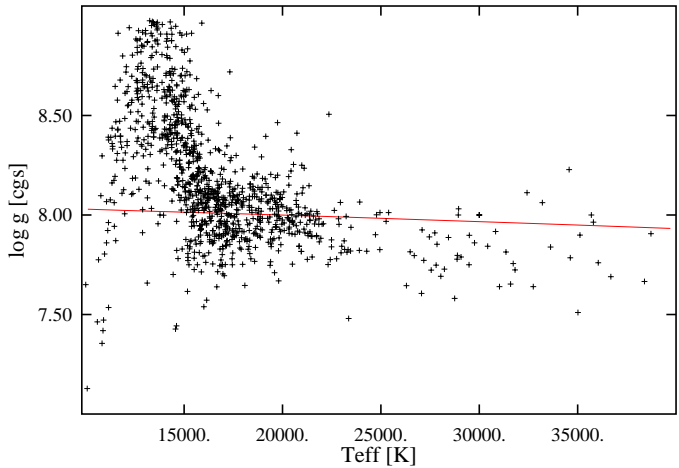}
\caption{Surface gravity \logg\ as a function of effective
  temperature. The continuous (red) line is the sequence for a
  constant mass of 0.6\,\Msun.
\label{figlogg_vs_teff}}
\end{figure}

\onllongtab{
\begin{longtab}
\begin{longtable}{rrrrrrrr}
\caption{\label{tabtg} Stellar parameters for the sample
  effective temperature (\Teff), surface gravity (\logg),
  S/N, distance (d), distance above Galactic plane (z), 
  and logarithmic abundances of H and Ca. Numbers in parentheses
  are formal errors from the fit; a zero means that \logg\ was kept
  fixed at 8.00 or, for the abundances, that these are upper
  limits.}\\ 
\hline\hline\\
SDSSJ & \multicolumn{1}{c}{\Teff} & \multicolumn{1}{c}{\logg} & S/N & d  & 
  z   & \multicolumn{1}{c}{[H/He]} & \multicolumn{1}{c}{[Ca/He]} \\ 
      & \multicolumn{1}{c}{[K]}  &  \multicolumn{1}{c}{[cgs]}&     
      &[pc]&[pc] & \\
\hline \endfirsthead\\
\caption{continued.}\\
\hline\\
SDSSJ & \multicolumn{1}{c}{\Teff} & \multicolumn{1}{c}{\logg} & S/N & d  & 
  z   & \multicolumn{1}{c}{[H/He]} & \multicolumn{1}{c}{[Ca/He]} \\ 
      & \multicolumn{1}{c}{[K]}  &  \multicolumn{1}{c}{[cgs]}&     
      &[pc]&[pc] \\
\hline\hline\\
\endhead
\hline
\endfoot
000116.49+000204.45&  10985 (130)&  8.000 (0.000)& 23.2& 220& 191&  -5.74 (0.00)& -10.62 (0.00)\\
000106.22+250330.00&  15194 ( 75)&  8.000 (0.000)& 15.4& 428& 254&  -4.73 (0.30)&  -7.37 (0.30)\\
000223.06+272358.50&  16886 ( 42)&  7.950 (0.028)& 45.1& 219& 123&  -5.48 (0.00)&  -6.56 (0.00)\\
000515.58+071313.71&  18650 ( 86)&  8.024 (0.033)& 30.5& 269& 217&  -4.70 (0.00)&  -5.09 (0.00)\\
000720.22-002325.40&  19276 (240)&  7.863 (0.068)& 14.9& 353& 309&  -4.21 (0.00)&  -4.97 (0.00)\\
000742.62+252422.50&  19916 (320)&  7.851 (0.085)& 11.0& 804& 477&  -3.85 (0.00)&  -4.36 (0.00)\\
000730.75+275111.90&  16048 (180)&  7.795 (0.132)& 10.1& 546& 305&  -4.84 (0.00)&  -6.36 (0.00)\\
000816.26+154609.44&  13547 ( 79)&  8.000 (0.000)& 18.4& 253& 181&  -5.50 (0.55)&  -8.68 (0.00)\\
\end{longtable}
\end{longtab}
}

\section{Atmospheric parameters}
Atmospheric parameters \Teff, \logg, abundances, or upper limits of
hydrogen and calcium were determined by a comparison of observations
with several grids of theoretical model atmospheres. The input physics
and procedures used are described in \cite{Koester10}. We use the
mixing-length approximation for convection in the version ML2
\citep{Fontaine.Villeneuve.ea81,Tassoul.Fontaine.ea90} with a mixing
length equal to 1.25 pressure scale heights. A pure helium grid -- for
technical reasons the logarithmic abundance ratio of hydrogen to
helium, $\log N_{\mathrm H}/N_\mathrm{He}$ (abbreviated [H/He]
henceforth) is -20.0 -- covers effective temperatures from 10\,000 to
50\,000\,K, with step widths ranging from 250 to 500\,K from low to high
effective temperatures. The surface gravity \logg\ ranged from 6.00 to
9.50, with step width 0.25. In addition to the pure helium grid above, we
also calculated grids with various [H/He] ratios: -6.0, -5.5, -5.0,
-4.5, -4.0, -3.5, and -3.0. These grids covered the same \Teff\ and
\logg\ values, except that the \logg\ range only covered  from 7.0 to
9.0. Details of the analysis are described in the following.

\subsection{Photometry}

From the synthetic spectra we calculated theoretical photometry in the SDSS
system. The effect of the surface gravity on the photometry is small; in the
first step of the fitting procedure, we thus kept this parameter fixed at the
canonical value of 8.0. A possible concern for the fainter objects at greater
distance is interstellar reddening, which leads to lower apparent
temperatures. We determined three different fit results: one assuming that the
reddening is negligible, a second that assumes the maximum reddening from the
\cite{Schlafly.Finkbeiner11} extinction map, and a third value that uses an
iterative procedure as described in \cite{Tremblay.Bergeron.ea11} and
\cite{Genest-Beaulieu.Bergeron14}. Basically, this method assumes that the
extinction is negligible within 100\,pc, and from there it increases linearly
to the maximum value at a vertical height $z=250$\,pc above the Galactic
plane. This approach also gives the best approximation for the distance and
$z$, although it should be kept in mind at this step a fixed radius
corresponding to \logg\ = 8.0 and a pure He atmosphere is assumed for all
objects.

\subsection{Spectroscopy} 
As a next step the observed spectra were fitted with the pure He grid, to
determine \Teff\ and \logg. A well known problem in many cases is that there
are two possible solutions corresponding to local $\chi^2$ minima, one below
and one above the region of \Teff\ = 24\,000 - 26\,000\,K, where the He lines
reach their maximum strength. The $\chi^2$ values of the two solutions are
usually very similar or even identical and cannot be trusted to select the
correct solution. We used the photometric fits, which do not have this
problem, as well as visual inspection of all fits, to minimize wrong choices.

Then, in all spectra we used an automatic measuring procedure to determine
\Halpha\ equivalent widths and uncertainties, or alternatively upper limits;
all positive and negative detections were confirmed by visual
inspection. These measurements were compared to theoretical equivalent widths
from the grids with various hydrogen traces as described above, and hydrogen
abundances [H/He], uncertainties, or upper limits were determined. With this
additional knowledge, the spectral fits were repeated with the grid that
matched to the measured abundance most closely, and this whole procedure was
iterated until the parameters were determined with (almost) fully consistent
theoretical models. As a final step, the photometric fit was repeated, but now
keeping all parameters from the spectroscopic results fixed and solving only
for the consistent distance and Galactic position. Table~\ref{tabtg} (online
only) contains the final results of this analysis.

\begin{table*}
\caption{ Average surface gravity \logg\ as a function of effective temperature
  \Teff\ in 19 intervals. Second column: average \logg\ and, in parentheses,
  1$\sigma$ width of the distribution. Third column: average mass, error of
  the average and 1$\sigma$ width of the distribution. N is the number of
  objects in the interval. BR is the number normalized with the cooling time
  through the interval and the luminosity to the 3/2 power. \label{tabresults}} 
% \small 
\centering
\begin{tabular}{cccrc}
\hline
  \Teff   &  \logg   &    $M$      &  N  & BR \\
  \  [K]    &  [cgs]   &    [\Msun]  &     &    \\
\hline
\noalign{\smallskip}
 10000 - 12000 & 8.166 (0.379) & 0.696 (0.029,0.208) &  52  &   2.092\\
 12000 - 13000 & 8.564 (0.234) & 0.928 (0.017,0.142) &  69  &   4.018\\
 13000 - 14000 & 8.577 (0.290) & 0.937 (0.019,0.173) &  83  &   3.928\\
 14000 - 14500 & 8.535 (0.236) & 0.914 (0.017,0.141) &  69  &   5.595\\
 14500 - 15000 & 8.384 (0.283) & 0.825 (0.018,0.165) &  81  &   6.125\\
 15000 - 15500 & 8.250 (0.276) & 0.741 (0.017,0.165) &  94  &   6.287\\
 15500 - 16000 & 8.137 (0.228) & 0.673 (0.015,0.135) &  84  &   5.284\\
 16000 - 16500 & 8.046 (0.175) & 0.621 (0.012,0.098) &  71  &   4.013\\
 16500 - 17000 & 8.021 (0.169) & 0.607 (0.011,0.095) &  72  &   3.865\\
 17000 - 17500 & 8.015 (0.171) & 0.605 (0.012,0.097) &  62  &   3.089\\
 17500 - 18000 & 8.014 (0.113) & 0.603 (0.009,0.062) &  44  &   2.093\\
 18000 - 19000 & 8.007 (0.128) & 0.601 (0.008,0.070) &  70  &   1.471\\
 19000 - 20000 & 8.011 (0.138) & 0.605 (0.009,0.077) &  65  &   1.165\\
 20000 - 22000 & 7.991 (0.119) & 0.597 (0.007,0.066) &  79  &   0.595\\
 22000 - 24000 & 7.898 (0.175) & 0.556 (0.019,0.095) &  24  &   0.161\\
 24000 - 26000 & 7.934 (0.079) & 0.574 (0.016,0.042) &   7  &   0.046\\
 26000 - 28000 & 7.789 (0.099) & 0.509 (0.014,0.045) &  11  &   0.074\\
 28000 - 30000 & 7.828 (0.132) & 0.532 (0.018,0.063) &  12  &   0.087\\
 30000 - 40000 & 7.841 (0.171) & 0.550 (0.016,0.079) &  23  &   0.063\\
\hline
\end{tabular}
\end{table*}

\section{Analysis}

\subsection{Surface gravity and masses}
Figure~\ref{figlogg_vs_teff} shows the surface gravity versus effective
temperature for all objects, except for a few low temperature DBs,
where the spectroscopic fitting did not converge on a \logg\ within
the grid; Table~\ref{tabtg} shows more detail in numerical form. Some
features of this result are immediately apparent: 

\begin{itemize} 

\item the \logg\ and masses are significantly higher below 16\,000~K than for
  the hotter objects. This effect has been observed before, e.g., by
  \cite{Kepler.Kleinman.ea07}, BW11, and \cite{Kepler.Pelisoli.ea15}. BW11
  tentatively conclude that the large mass spread might be real for $\Teff >
  13\,000$~K, given the presence of larger and smaller \logg\ determinations
  in the same temperature interval and a comparison of spectroscopic and
  parallax distances of 11 objects.  Only one of the 11 objects has a really
  large spectroscopic mass, and for this one the two distances show a large
  discrepancy; the authors admit that at these low temperatures the limit of
  the spectroscopic method may have been reached, because of the weakness of
  the He lines.  We believe that the systematic change in our sample
  (Fig.~\ref{figlogg_vs_teff}) is caused by an imperfect implementation of
  line broadening by neutral helium, which dominates the broadening below
  16\,000\,K.

\item There is also a significant increase in the width of the
  \logg\ distribution, in addition to this systematic effect. The explanation
  is most likely the decreasing strength of the helium lines in conjunction
  with the moderate resolution and S/N of the SDSS spectra, which increase the
  errors of the individual parameter determinations. Because of these results,
  we only use the solution with \logg\ fixed at 8.0 for $\Teff < 16\,000$\,K
  and assume an error for \logg\ of 0.25.

\item The average mass in Table~\ref{tabresults} shows little variation in the
  range $16\,000 \leq \Teff \leq 22\,000$\,K with an average over this whole
  range of $0.606 \pm 0.004$\,\Msun. In six of the seven highly populated
  intervals, the average agrees with this value within the $1\sigma$ errors,
  and in the other within $2 \sigma$, which is close to expectation, if the
  average mass is indeed constant. We conclude that this is a realistic
  estimate of the average mass of DB (including DBA, DBZ) white dwarfs and
  that any averaging that includes cooler or hotter objects will necessarily
  lead to erroneous results. For example, for our complete sample we get an
  unreliable average of $0.706 \pm 0.006$\,\Msun.

\item DBs at 10\,000\,K are approximately $5\times 10^8$ years older than at
  30\,000\,K. They may originate from more massive progenitor stars on
  the main sequence and end up as higher mass white dwarfs. Using data from
  \cite{Salaris.Serenelli.ea09} and \cite{Pietrinferni.Cassisi.ea04}, we
  estimate that this effect could account for a difference of about
  0.05\,\Msun\ over the observed range. This could explain the slightly lower
  mass at the high \Teff\ end, but certainly not the large increase below
  16\,000\,K.

\item Both Fig.~\ref{figlogg_vs_teff} and Table~\ref{tabresults} show a
  deficiency of objects -- almost a gap -- in the interval 24\,000-26\,000\,K,
  in comparison to both neighboring intervals. To demonstrate that such a gap
  is not expected from evolution or observational effects, we calculated the
  quantity BR in the last column of the table. This is the number of objects
  divided by the cooling time during the interval and by the luminosity
  $L^{3/2}$, and scaled by an arbitrary constant for easier comparison. The
  first factor takes the different times spent in each interval during the
  cooling into account, and the second the larger observation volume for
  intrinsically brighter objects. If the observations were complete and a
  magnitude-limited sample, this number would be proportional to the birthrate
  of DB white dwarfs, and thus be a constant, if all DBs originate at high
  \Teff\ and only evolve through the range of our sample (see
  below). Obviously these conditions are not fulfilled; nevertheless, even
  with these numbers, the gap 24\,000-26\,000\,K is significant. (We note that
  the apparent gap is centered on 26\,000\,K, and the argument would be even
  more convincing had we used the interval 25\,000--27\,000\,K.)\\
\end{itemize}

Our explanation of the last point is as follows: This temperature region is
exactly where the helium lines have their maximum strength as a function of
effective temperature. If the predicted line strengths are greater than
attained by the observed objects, then the fitting procedure will force them
to a solution on either side of the maximum. BW11 also discuss this
possibility and conclude that a mixing-length parameter $\alpha = 1.25$ (which
is also our choice) provides the smoothest distribution of stars over the
temperature range. However, because of their smaller sample, there are only
$\sim$20 objects between 20\,000 and 30\,000~K, compared to 134 in our
sample. Judging by Fig.\,2 in BW11, a slightly larger $\alpha$ would probably
give a more plausible distribution for our sample.

An alternative suggestion, made by the referee, is a possible small
error in the SDSS flux calibration (see below for further
discussion). A possible indication for this is that two of the common
objects between the BW11 and our sample have \Teff\, $\sim$26\,000\,K
in BW11, but 24\,000 and 28\,000\,K in our determination.

\begin{figure}
\centering 
\includegraphics[width=0.49\textwidth]{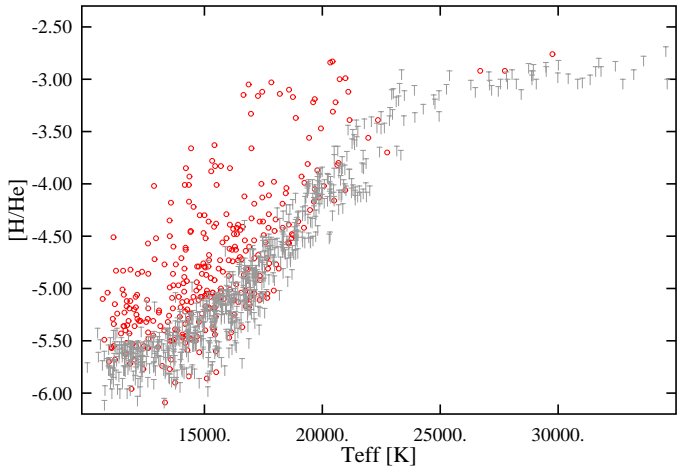}
\caption{Logarithmic hydrogen abundance [H/He] as a function of \Teff. Red
  circles indicate detected abundances, black symbols indicate upper limits.
\label{figH_vs_teff}}
\end{figure}

\begin{figure} \centering
\includegraphics[width=0.49\textwidth]{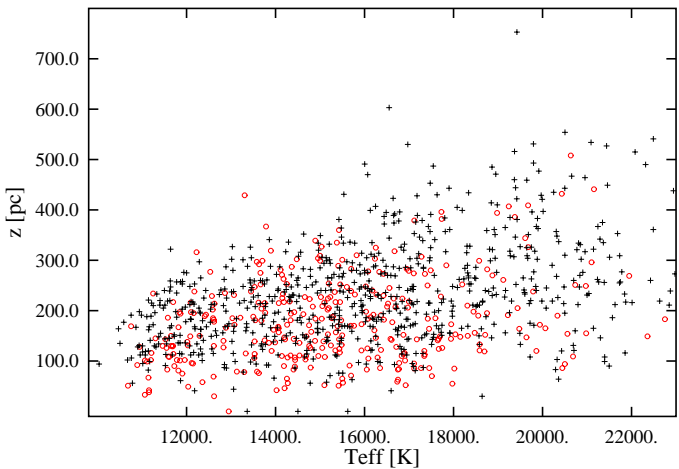}
\caption{Distribution of DBAs (red circles) and DBs (black crosses)
with height $z$ above the Galactic plane. The general decrease of
distances towards lower \Teff\ is very likely caused by the lower
luminosity of the cooler objects.  \label{figzH_vs_Teff}} 
\end{figure}

\begin{figure}
\centering 
\includegraphics[width=0.49\textwidth]{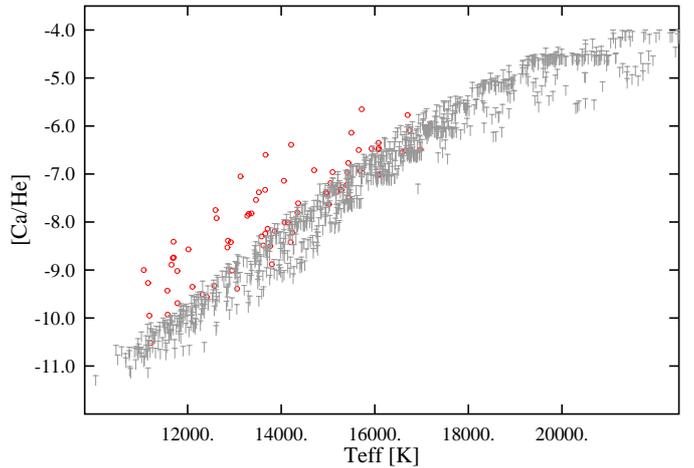}
\caption{Logarithmic calcium abundance [Ca/He] as a function of \Teff. Red
  circles represent detected abundances, and black symbols the upper limits.
\label{figZ_vs_Teff}}
\end{figure}

\subsection{Hydrogen abundance}
The hydrogen abundance as a function of \Teff\ is shown in
Fig~\ref{figH_vs_teff}. The detection limit is high at high temperatures
because the hydrogen spectral lines get weaker. On the other hand, at low
temperatures, much smaller abundances can easily be detected.  The abundances,
however, do not reach the same level as at the high \Teff\ end, because the
hydrogen gets diluted by the increasing depth of the convection zone.  Out of
a total of 1036 objects with $\Teff < 23\,000$\,K, 329 or 32\% show
hydrogen. However, our sample extends over a large range of S/N values and
positions in the Galaxy that have an influence on the observed DBA
fraction. This is demonstrated in Table~\ref{tabhsnz}.

\begin{table}
\caption{Dependence of the DBA/(DBA+DB) number ratio on the signal-to-noise
  (S/N) and position in the Galaxy $z$. The statistics use only objects with
  $\Teff \leq 23\,000$\,K to minimize observational bias that is due to the
  low detection probability at higher \Teff.
 \label{tabhsnz}} 
\small \centering
\begin{tabular}{rrrr}
\hline
\noalign{\smallskip}
  S/N, $z$[pc]   &  N(DBA)  &   N(DB)   & \%DBA\\
\hline
\noalign{\smallskip}
      S/N $\geq$ 10, $z \leq 250$ & 267& 449& 37.3\\
      S/N $\geq$ 10, $z >  250$ &  62& 258& 19.4\\
      S/N $\geq$ 20, $z \leq 250$ & 191& 179& 51.6\\
      S/N $\geq$ 20, $z >  250$ &  12&  27& 30.8\\
      S/N $\geq$ 30, $z \leq 250$ & 106&  55& 65.8\\
      S/N $\geq$ 30, $z >  250$ &   0&   4&  0.0\\
      S/N $\geq$ 40, $z \leq 250$ &  46&  15& 75.4\\
      S/N $\geq$ 40, $z >  250$ &   0&   0&  0.0\\
      S/N 10 - 15, $z \leq 250$ &  29& 140& 17.2\\
      S/N 10 - 15, $z >  250$ &  25& 140& 15.2\\
\hline
\end{tabular}
\end{table}

\begin{table}
\caption{Similar to Table~\ref{tabhsnz} but for Ca:
dependence of the DBZ/(DBZ+DB) number ratio on the
  S/N and position in the Galaxy $z$. The
  statistics use only objects with $\Teff \leq 17\,000$\,K.
 \label{tabcsnz}} 
\small \centering
\begin{tabular}{rrrr}
\hline
\noalign{\smallskip}
  S/N, $z$[pc]   &  N(DBZ)  &   N(DB)   & \%DBZ\\
\hline
\noalign{\smallskip}
      S/N $\geq$ 10, $z \leq 250$ &  62& 484& 11.4\\
      S/N $\geq$ 10, $z >  250$ &  13& 144&  8.3\\
      S/N $\geq$ 20, $z \leq 250$ &  34& 233& 12.7\\
      S/N $\geq$ 20, $z >  250$ &   0&  11&  0.0\\
      S/N $\geq$ 30, $z \leq 250$ &  14& 100& 12.3\\
      S/N $\geq$ 30, $z >  250$ &   0&   0&     \\
      S/N $\geq$ 40, $z \leq 250$ &   4&  41&  8.9\\
      S/N $\geq$ 40, $z >  250$ &   0&   0&     \\
      S/N 10 - 15, $z \leq 250$ &  12& 124&  8.8\\
      S/N 10 - 15, $z >  250$ &  10&  82& 10.9\\
\hline

\end{tabular}
\end{table}

The S/N ratio is an obvious factor influencing the
detectability of hydrogen. The percentage of DBAs increases very
significantly with increasing S/N, from 37\% for the subsample with $z
\leq 250$\,pc to 75\% at the highest S/N. This is even higher than the
values of 44\% found by BW11 and 55\% by \cite{Voss.Koester.ea07},
regarded as being lower limits in those papers. Our findings suggest
that practically all DBs  show some trace of hydrogen if the
resolution and S/N are high enough.

With the large size of our sample and the excellent SDSS photometry, for the
first time we can try to study the position of DBs and DBAs in the Galaxy, in
particular the height above the Galactic plane. This is an important quantity
for the question: is the hydrogen content due to external processes, e.g.,
accretion from interstellar gas, or from the remnants of a planetary system,
or to an intrinsic process like dilution of an outer hydrogen envelope in a
developing helium convection zone.  Figure~\ref{figzH_vs_Teff} shows this
distribution, which demonstrates a concentration of the DBAs towards the
Galactic plane. The effect becomes even clearer in Table~\ref{tabhsnz}, which
shows a decrease in the DBA percentage with height $z$. Before jumping to
premature conclusions, however, we need to take into account that the group
with $z > 250$\,pc has on average larger distances and lower S/N, which
possibly alone could explain the $z$ dependence. To test this we have taken a
progressively smaller limit on the S/N, to make both groups more
comparable. Taking only objects with {10 $\leq$ S/N $\leq$ 15,} the average
S/N and its distribution become almost identical. For this sample the
difference between the two groups disappears (last lines in
Table~\ref{tabhsnz}): taking samples with comparable S/N, there is no obvious
concentration toward the Galactic plane.

\begin{figure}
\centering 
\includegraphics[width=0.49\textwidth]{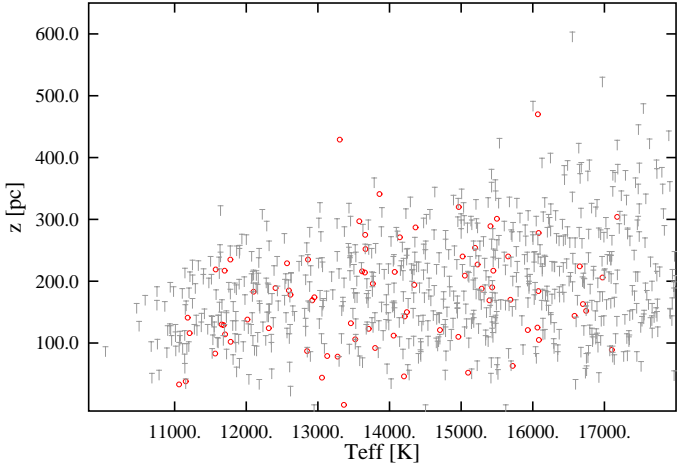}
\caption{Distribution of DBZs (red circles) and DBs (black limit
  symbols) with height $z$ above the Galactic plane.
\label{figzZ_vs_Teff}}
\end{figure}

\begin{figure}
\centering 
\includegraphics[width=0.49\textwidth]{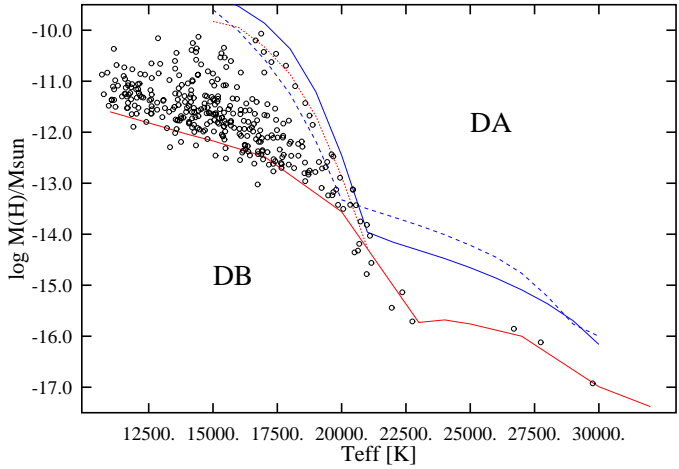}
\caption{Total hydrogen mass in the convection zone. The continuous red line
  is the transformed lower limit to the positive detections in
  Fig.~\ref{figH_vs_teff}. The dotted red curve is the expected location for
  an abundance [H/He] = -3, which coincides with the continuous red curve at
  high temperatures. The blue curves indicate the expected hydrogen masses for
  abundances of -2 (continuous) or -1 (dotted).
\label{fighydmass}}
\end{figure}

\begin{figure}
\centering 
\includegraphics[width=0.49\textwidth]{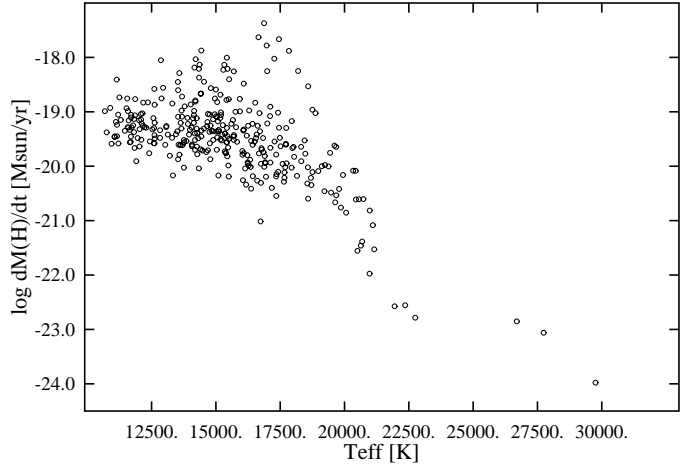}
\caption{Average rate for the increase in the total hydrogen mass with
  age of the white dwarf.
\label{fighydrate}}
\end{figure}

\begin{figure}
\centering 
\includegraphics[width=0.49\textwidth]{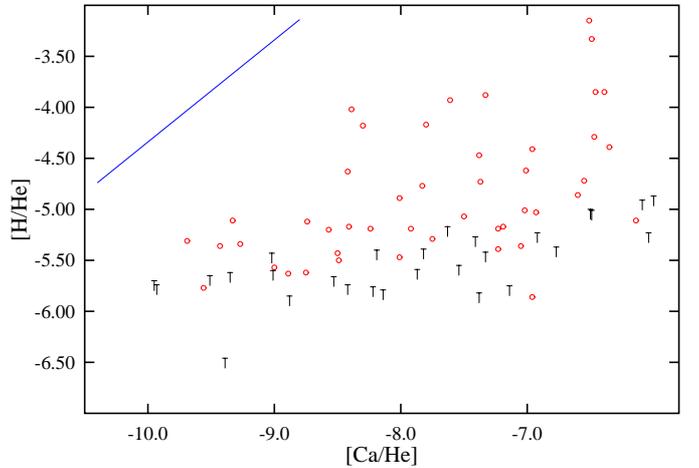}
\caption{Hydrogen abundance or upper limit versus Ca abundance in all
  DBZs. The [H/Ca] ratio is smaller than the solar value (continuous
  line) in all objects.
\label{figH_vs_Ca}}
\end{figure}

\subsection{Metals in DBs: the DBZs}
Seventy-five objects show the H+K resonance lines of ionized calcium -- this
is the only metal detected in our sample.  Figures~\ref{figZ_vs_Teff} and
\ref{figzZ_vs_Teff} and Table~\ref{tabcsnz} show the distribution of the DBZs
with \Teff, S/N, and $z$ in the same way as for hydrogen in the previous
figures and tables. Surprisingly, the fraction of DBZs shows little variation
with the S/N ratio, which might even be explainable by the smaller
numbers. The only explanation we can find is that the \ion{Ca}{ii} lines are
much narrower and deeper than \Halpha, and therefore easier to detect, even at
low S/N.  The change of the ratio with $z$ is smaller than that for hydrogen
(from 11.4 to 8.3\%) and even reverses when using the limited sample with S/N
between ten and 15. The numbers are still relatively small, but we can
definitely state that they do not prove any preference for accretion of
interstellar matter close to the Galactic plane (Fig.~\ref{figzZ_vs_Teff}).

\subsection{Error estimates} 
The errors for \Teff\ and \logg\ in Table~\ref{tabresults} are only formal
statistical errors and thus underestimate the real uncertainties. Fortunately
we have multiple spectra for 149 objects, leading to 162 pairs of
\Teff\ determinations and 72 for \logg\ (excluding those with fixed 8.0). The
average absolute differences are 3.1\% for \Teff, 0.12 for \logg, 0.18 for
[H/He], and 0.25 for [Ca/He]. Since the models and fitting procedures are
exactly the same, these are uncertainties due to the observations caused, for
example, by different S/N or reductions.

There are 27 objects in common with the sample of BW11. If we compare only the
17 DBs with $\Teff \geq 16\,000$\,K, where we determine both parameters, the
systematic difference for \Teff\ is -1.3\%, i.e. our \Teff\ are slightly
larger on average). The dispersion is 4.6\%, a reasonable number given
that our internal uncertainties are already 3.1\%. The differences in surface
gravity are larger, with a systematic effect of 0.095\,dex. In their study of
DA white dwarfs, \cite{Genest-Beaulieu.Bergeron14},
\cite{Tremblay.Bergeron.ea11}, and \cite{Gianninas.Bergeron.ea11} find similar
differences between their own spectra and SDSS samples. They conclude that the
most likely reason is a small residual calibration error of the SDSS
spectra. It is plausible that such an error would also affect the DB
spectra. The \logg\ dispersion between the BW11 and our results for \logg\ is
0.073\,dex. The larger internal error that we obtained above may be influenced
by different S/N values for the multiple spectra of the same object.

\section{Results and discussion}
The overall features of our DB sample are similar to the results of
BW11 and \cite{Voss.Koester.ea07}: an apparent increase in the masses
toward lower temperatures, mean mass around 0.6\,\Msun, several DBs
within the so-called DB gap above 30\,000\,K, and a large number of DBs
contaminated with hydrogen abundances [H/He] between -6 and
-3. Apparent differences may at least be partly attributed to the
tenfold increase of the sample size.

We do not find the apparent dichotomy at the very cool end between normal
masses and a few very high masses near 1.2\,\Msun, which probably motivated
BW11 to exclude only the few objects in this range when determining a mean
mass for DBs. In our sample there is clearly a continuous distribution; the
increase in masses starts below 16\,000\,K. Since the convection zone develops
near 30\,000\,K and deepens significantly below 20\,000\,K, we think that
neutral broadening is a more likely explanation than the convection theory,
which is thought to be the reason behind a similar effect in the DAs
\citep{Tremblay.Ludwig.ea13}. For the averaging, we thus use the intervals
between 16\,000 and 22\,000\,K, with many objects and a constant mass
throughout, which results in a lower mean mass of $0.606 \pm
0.004$\,\Msun\ compared to 0.671\,\Msun\ in BW11. The mean masses of
\cite{Voss.Koester.ea07} agree approximately with our current result, but
since different mixing-length parameters and different intervals for the
averaging were used, the comparison is not very meaningful. We emphasize again
that, as long as the reason for the \logg\ increase at the low \Teff\ end is
not understood, it is very important to only compare mean masses for
well-defined \Teff\ intervals. If we use our complete sample, we obtain a mean
mass of 0.706\,\Msun, which is even higher than the BW11 result. Our preferred
result agrees with the most recent determination of 0.603$\pm$0.002 for DA
white dwarfs from DR12 by Kepler et al. (2015a), but the caveat
above also applies to the DA vs. DB comparison.

\subsection{Hydrogen}
Our result for the highest S/N objects indicates that at least 75\% -- and
perhaps all DBs -- show some contamination with hydrogen. The difference
between DBs and DBAs is very likely just a question of the quality of the
observations and we no longer distinguish the mean masses for the two
groups. Since hydrogen lines are detected much more easily at low \Teff\,, the
detections and upper limits go down to abundances around -6. High abundances,
as found at high \Teff\,, are not detected in this range, although they would
be found easily. The obvious reason is the deepening of the convection zone
below $\sim$18\,000\,K, and indeed, if multiplied with the mass in the
convection zone, the total hydrogen masses seem to increase towards lower
\Teff\ (Fig.~\ref{fighydmass}). As a reminder, all hydrogen originally present
or accreted from any source should always stay at the top of the envelope, in
this case within the outer helium convection zone. The total hydrogen mass in
an object can thus never decrease with time.

Continuous accretion from the interstellar medium could explain such
an increase; dividing the hydrogen masses by the cooling age of the
white dwarf indicates time-averaged accretion rates of $10^{-17}$ to
$10^{-24}$\,\Msun/yr (Fig.~\ref{fighydrate}). The size of these
average rates is  reasonable \citep{Dupuis.Fontaine.ea93}, 
although the huge spread and the increase toward lower temperatures
might be difficult to explain. An alternative proposal, the continuous
accretion of comets from an Oort cloud \citep{Veras.Shannon.ea14},
could also explain such an increase toward older white dwarfs, but
this faces the same problems.

To put these results into proper perspective we have to consider
the strong observational selection effects apparent in
Fig.~\ref{figH_vs_teff}. Transforming the approximate location of the
observable lower limits for the detected DBAs into lower limits for the total
H mass in the convection zone leads to the continuous red line in
Fig.~\ref{fighydmass}; because of the spread in S/N and range in
\logg,\ there is a transition region and some objects are still found below
our chosen curve. Below the red line we do not expect to find hydrogen in the
objects of our sample. The region is of course not empty but filled with DBs
with upper limits to the hydrogen mass between $10^{-16}$ and
$10^{-12}$\,\Msun, depending on \Teff.

At the hot end all abundances are [H/He] $\approx  -3.0$; we have
continued the (theoretical) location for this abundance with the
dotted red curve toward lower \Teff. Almost all of our objects are
confined to the region between these two red curves. Larger abundances
are not found, but with our model calculations we can   predict their
location in Fig.~\ref{fighydmass}. For abundances larger than
 {$\sim-2$}
the convection zones in the atmosphere become tiny
or absent. Such models are not realistic because the hydrogen would
diffuse upward immediately and turn the star into a hydrogen-rich DA
-- and thus out of our sample. In the upper right part we do not
expect to find any DBAs as this region is occupied by DAs. The upper left part
also seems sparsely populated. Rare objects like GD16 could fit in
there with 11\,000\,K and $\log M_{\mathrm H}/\Msun \approx-9$
\citep[][Gentile-Fusillo et al. in prep]{Koester.Napiwotzki.ea05}, but
these can easily be misclassified as a DA.

In summary: we find exactly those objects that can exist in our sample, given
the observational constraints and the physics of gravitational settling. The
morphology of Fig.~\ref{fighydrate} is very similar to Fig.~\ref{fighydmass},
because the cooling ages change only by 1.5 dex over the whole range of the
figure, and the remarks concerning the observable objects are also applicable
to the hypothetical accretion rates. The conclusion from this discussion is
that Figs.~\ref{fighydmass} and \ref{fighydrate} do not prove that there is
any increase in the total hydrogen mass with time. The descendants of the hot
DBAs with $M_{\mathrm H}/\Msun = 10^{-16}$ are not the observed cool DBAs but
are lost from sight when the H abundance goes below the observable limit.
This has to be taken into account when trying to derive conclusions about the
origin of hydrogen in DBs from our present results.

Although our approach and presentation are completely different from
the theoretical calculations of \cite{Macdonald.Vennes91}, we agree
with their prediction that helium-rich DBs with a hydrogen mass $
>10^{-14}$\,\Msun\ can only appear below $20\,000 - 22\,000$\,K,
depending slightly on the version of the mixing-length theory
applied. According to their Fig.\,1 and Table\,1, higher hydrogen
masses appear at progressively lower \Teff, where they change from a
pure hydrogen object into a DBA (or possibly a DB). If we look closely at the
details, however, significant differences appear. To take a specific
example: a DA with $M_{\mathrm H}/\Msun = 10^{-12}$ should turn into a
DBA only at 11,300\,K, whereas we already find such objects  around
19\,000\,K. These problems led BW11 to the hypothesis that hydrogen
might not be completely mixed within the He convection zone, but
floating only near the top. This would decrease the total amount of H
present for a given abundance and effective temperature, thus allowing
for a dredge-up at higher \Teff.  We consider this scenario unlikely
since the convection velocities in the highly turbulent convection
zone are many orders of magnitude larger than the diffusion velocities
of hydrogen in helium. Instead, we prefer to speculate that the
discrepancies are due to our imperfect description of convection with
the mixing-length theory. Nature somehow seems to  manage a complete
mixing at 20\,000\,K, although theory predicts it only happens at 11\,300\,K.

Does this mean that the origin of hydrogen in DBAs is the mixing of a residual
small hydrogen layer in the range 30\,000 - 40\,000\,K, which is left over
from the previous evolution?  In the framework of interstellar accretion as a
source of the hydrogen, one would have to assume a typical average hydrogen
accretion rate of $10^{-19}$\,\Msun/yr for the majority of the cool
objects. In $10^5$ years the star would accumulate an outer H layer of
$10^{-14}$\,\Msun, enough to become a DA. This is a very short time compared
to the cooling age at 30\,000\,K, and for the later evolution it makes no
essential difference if the DA arrives with a thin evolutionary H layer or
acquires it almost instantaneously through accretion. Both scenarios require
that a significant fraction of DA white dwarfs near \Teff\ = 30\,000\,K have
thin hydrogen envelopes in the range of $10^{-16} - 10^{-10}$\, \Msun. It also
requires that many DB/DBA white dwarfs are only borne around 20\,000\,K and,
thus, that their space density, corrected for cooling times (i.e., the
luminosity function), is smaller at higher \Teff. This is exactly what was
found by BW11: the DB to DA ratio of all white dwarfs in the Palomar-Green
sample increases sharply around 20\,000\,K during the cooling sequence.

A decision about the origin of hydrogen then boils down to the
question of whether it is more likely for stars  to have environments at
30\,000\,K that lead to accretion rates differing by six orders of
magnitude or to have residual hydrogen layers from the previous
evolution between $10^{-16}$ and $10^{-10}$\,\Msun. In view of the
absence of any correlation between the occurrence of the DBAs and height
above the Galactic plane, we favor the second alternative.

As a result of the existence of cool DBs without any visible hydrogen, BW11
concluded that there must be two channels for DBs. One channel consists of DAs
that are transformed into DBs through convective mixing, and the other of DBs
that never change during evolution. If we accept the view that the hydrogen is
residual hydrogen from the previous evolution, then a more natural explanation
would be that the hydrogen mass (in DBs) extends from $\sim
10^{-10}$\,\Msun\ not just down to $10^{-16}$\,\Msun,\ but even lower. The
very existence of DBs in the so-called DB gap that exists between 30\,000 and
45\,000\,K proves this. In connection with this it is interesting that the
hydrogen layer masses in the ZZ Ceti (DA!) white dwarfs, which are determined
by pulsational properties \citep{Romero.Corsico.ea12, Castanheira.Kepler09},
cover the range from $\sim 10^{-4}$ to a lower limit of $\sim
10^{-10}$\,\Msun. It is tempting to wonder if we really need two different
scenarios for the origin of the hydrogen-rich versus the helium-rich white
dwarf sequences or whether have to accept a continuum of hydrogen layer
thickness from the canonical DA value of $10^{-4}$\,\Msun\ down to zero, but
such considerations are beyond the scope of this work.

\begin{figure}
\centering 
\includegraphics[width=0.49\textwidth]{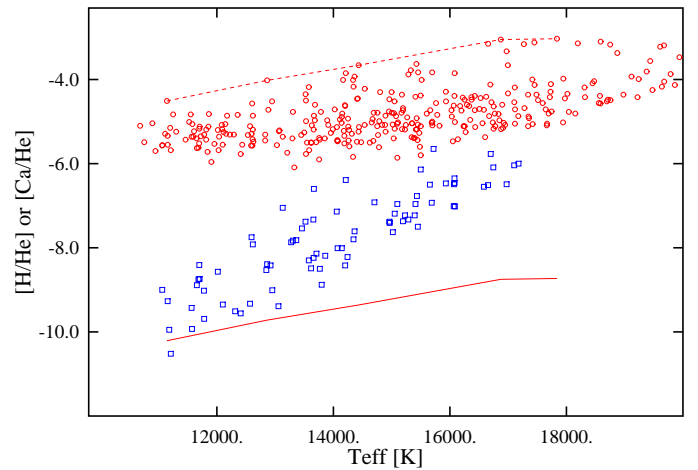}
\caption{Hydrogen (red circles) and Ca (blue squares) abundance versus
  \Teff. The   red dotted  line is an empirical upper limit to H abundances,
  drawn through those objects with the highest abundances. The
  continuous line is shifted downward by 5.7 dex, the solar H/Ca
  ratio. Stars with a solar or higher H/Ca would be found below this
  line, which is, however, below the visibility limit of Ca for all but
  the coolest stars.
\label{figH_and_Ca}}
\end{figure}

\subsection{Calcium}
Unlike hydrogen, calcium does not accumulate in the outer layers with time but
diffuses downward with a timescale that is much shorter than the cooling
timescale. The origin therefore must be external. If accretion from the ISM
(gas and dust) were the source for both hydrogen and calcium, we would thus
expect that the Ca/H ratio is always smaller than the solar (ISM) value. 
  On the contrary, Fig.~\ref{figH_vs_Ca} shows the opposite: the ratio in all
objects is at least a factor of ten larger than solar. This is a well known
fact in many metal-polluted white dwarfs and has always been a problem for the
ISM accretion hypothesis. Again, however, selection effects are important in
this respect and may lead to wrong conclusions: Fig.~\ref{figH_and_Ca}
demonstrates that there is only a very slim chance of finding such objects at
the lowest \Teff\ with the solar ratio in our sample. They must exist, since
eventually the calcium will completely diffuse out of the atmosphere, but at
this time we cannot find them any more. That we find many objects with Ca is
significant and means that, at least in all of these objects, the accretion is
currently of extremely hydrogen-poor material.  Together with the missing
correlation between DBZs and Galactic position, this adds more weight to the
hypothesis that ISM accretion cannot explain the observed facts, which leaves
the currently favored accretion from a circumstellar dust disk as remnant of a
planetary system as the only viable scenario.

\begin{figure}
\centering 
\includegraphics[width=0.49\textwidth]{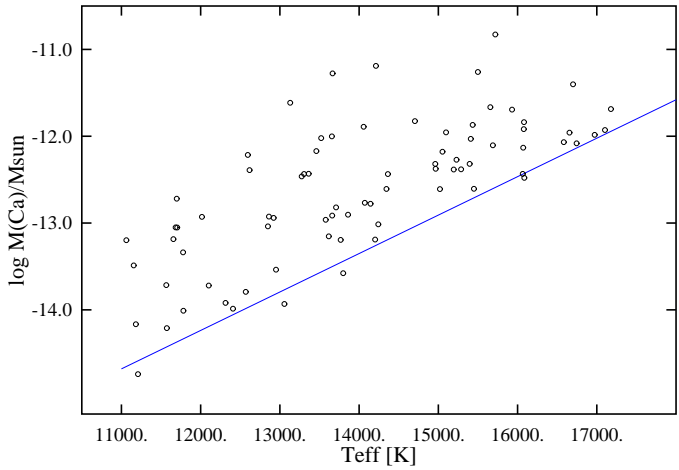}
\caption{Total calcium mass within the convection zone. The blue
  continuous line indicates the approximate lower limits, transformed
  from Fig.~\ref{figZ_vs_Teff}.
\label{figcamass}}
\end{figure}

\begin{figure}
\centering 
\includegraphics[width=0.49\textwidth]{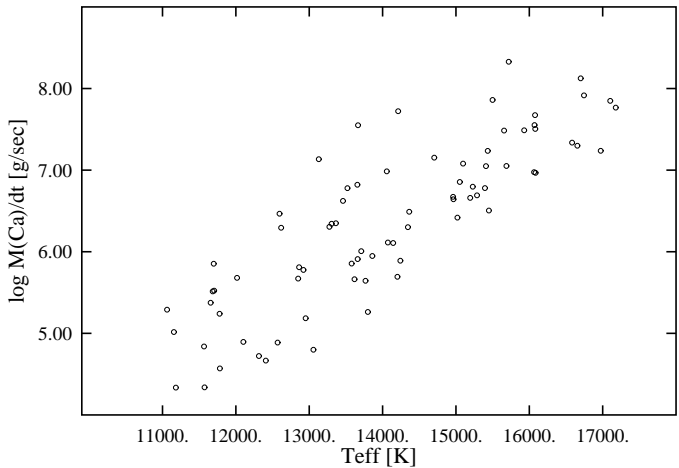}
\caption{Accretion rate of Ca in g/s, assuming equilibrium between
  accretion and diffusion.
\label{figcadiff}}
\end{figure}

The total mass of calcium within the convection zone (Fig.~\ref{figcamass})
shows the opposite behavior from that of hydrogen -- the total amount
decreases significantly toward lower temperatures, confirming our assumption
that no accumulation occurs with time. It is thus not meaningful to calculate
average rates as done for hydrogen, but instead the important timescale for
metals would be the diffusion timescale. Assuming equilibrium between
diffusion and accretion, we can calculate the calcium accretion flux
\citep{Koester09}. This is presented in Fig.~\ref{figcadiff}, where we have
changed the accretion flux unit to g/s, the preferred choice in the case of
metal polluted white dwarfs. The lower limit is obtained from the lower limit
to the observed abundances in Fig.~\ref{figZ_vs_Teff}. The strong decline with
decreasing \Teff\ is not easy to understand in the currently favored
explanation of accretion from a circumstellar disk. If indeed most of the
objects are in an equilibrium state between diffusion and accretion, we would
expect, on average, constant accretion rates from high to low temperatures, as
has been found for metal pollution in DAs \citep[see Fig.~8 in][]
{Koester.Gaensicke.ea14}. The problem is already visible in the abundance
distribution of Fig.~\ref{figZ_vs_Teff}; the change of 4.5 orders of magnitude
is much more than could be explained by dilution in the He convection zone,
which changes only by 1.6 orders over this range. This leads us to ask how
accurate our calculations of the convection zone depth are.

\subsection{The depth of the helium convection zone} 
Our envelope code starts at some specific position in the atmosphere model
(usually $\tau_\mathrm{Ross} = 100$) and integrates the stellar structure
equations inward. The equation of state (EOS) used for hydrogen and helium
models is that of \cite{Saumon.Chabrier.ea95}. For a model with \Teff\ =
10\,000\,K, \logg\ = 8.0 the base of the convection zone is reached at density
(g\,cm$^{-3}$) $\log \rho = 2.76$ and temperature (K) $\log T = 6.56$. The
zone runs through the region of pressure ionization of He (see Fig.~2 of the
paper cited above), which is not treated explicitly but bridged by an
interpolation scheme. The most important quantity for the envelope structure
is the adiabatic gradient, since the convection zone is very nearly
adiabatic. Looking at Fig.~23 in \cite{Saumon.Chabrier.ea95}, it is obvious
that different EOS give different results for the adiabatic
gradient. Considerable scatter is also apparent, which is probably caused by
the numerical calculation of second derivatives in the free-energy
minimization procedure. As a test of the sensitivity of our envelope
structure, we have made a calculation with the adiabatic gradient artificially
set to 0.4 everywhere. This decreases the mass in the convection zone by
almost three orders of magnitude. While this is certainly an extreme
assumption, the extent of the convection zone, and with it the derived total
masses of hydrogen and calcium, as well as the diffusion timescales, depend
very sensitively on details of the EOS used and could be uncertain by large
factors.

\section{Conclusions}
We analyzed photometry and spectra of the largest sample of
helium-rich stars studied so far. The estimated masses show a
significant increase below \Teff = 16\,000\,K, which we attributed to
imperfect implementation of line broadening by neutral helium
atoms. Using  the range from 16\,000 to 22\,000, where the average
mass does not change, do we find an average of $0.606\pm0.004$, which is
identical to the latest determination for DAs from DR12 (Kepler et al. 2015a).

At least 75\% of the helium-rich stars show contamination with hydrogen. The
total amount of hydrogen is larger at low \Teff\ than at the hot end. However,
we show that when selection effects are taken into account this does not prove
that the hydrogen in any single object increases with time. The unavoidable
conclusion is that a significant number of DAs must appear at 30\,000\,K with
thin hydrogen layers. Whether (i) these are always $\sim 10^{-16}$\,\Msun\ and
then increase during evolution by average accretion rates, which have to span
the range from $10^{-23} - 10^{-17}$\,\Msun/yr, or whether (ii) the
evolutionary hydrogen layer mass spans the range from $10^{-17} -
10^{-10}$\,\Msun\ and further accretion is unimportant cannot be distinguished
from the current data. Given that there is no correlation of the DBA numbers
with distance above the Galactic plane and also that the observed metals
cannot be explained by ISM accretion, we prefer the thin H-layer
alternative. We admit that the current theoretical calculations predict the
existence of stable DBA models with $M_\mathrm{H}$ from $10^{-14} -
10^{-10}$\,\Msun\ between 15\,000 and 21\,000\,K, but not any evolutionary
path that would lead there. The solution to this puzzle might be a more
physically sound treatment of convection and convective mixing.

About 10-12\% of the DBs are contaminated by traces of calcium. This can only
be accreted from an external source, and it is very clear that, in all cases,
the accreted matter is extremely hydrogen-poor. ISM accretion with solar
abundances can be ruled out, and the currently favored model of accretion from
a dust disk is supported. If we assume equilibrium between accretion and
  downward diffusion at the bottom of the convection zone we find calcium
accretion rates that decrease strongly with decreasing \Teff. We have
currently no explanation for this and it might indicate that the extreme
physical conditions in the pressure ionization region of helium are not yet
adequately described.

\begin{acknowledgements} DK gratefully acknowledges support from the program
  Science without Borders, MCIT/MEC-Brazil, which provided the
  opportunity for extended visits to Porto Alegre.

Funding for SDSS-III has been provided by the Alfred P. Sloan Foundation, the
Participating Institutions, the National Science Foundation, and the
U.S. Department of Energy Office of Science. The SDSS-III web site is
http://www.sdss3.org/.  SDSS-III is managed by the Astrophysical Research
Consortium for the Participating Institutions of the SDSS-III Collaboration
including the University of Arizona, the Brazilian Participation Group,
Brookhaven National Laboratory, Carnegie Mellon University, University of
Florida, the French Participation Group, the German Participation Group,
Harvard University, the Instituto de Astrofisica de Canarias, the Michigan
State/Notre Dame/JINA Participation Group, Johns Hopkins University, Lawrence
Berkeley National Laboratory, Max Planck Institute for Astrophysics, Max
Planck Institute for Extraterrestrial Physics, New Mexico State University,
New York University, Ohio State University, Pennsylvania State University,
University of Portsmouth, Princeton University, the Spanish Participation
Group, University of Tokyo, University of Utah, Vanderbilt University,
University of Virginia, University of Washington, and Yale University.
\end{acknowledgements}

\end{document}